\documentclass{moriond}

\usepackage{amsmath,amssymb}

\def\be{\begin{equation}}
\def\ee{\end{equation}}
\def\bea{\begin{eqnarray}}
\def\eea{\end{eqnarray}}

\begin{document}
\vspace*{4cm}
\title{EFFECTS OF DENSE MATTER ON THE TIDAL DEFORMATIONS OF BINARY NEUTRON STAR INSPIRALS AND GRAVITATIONAL WAVES}

\author{ L. PEROT and N. CHAMEL }

\address{Institute of Astronomy and Astrophysics, \\ Universit\'e Libre de Bruxelles, CP 226,\\ Boulevard du Triomphe, B-1050 Brussels, Belgium}

\maketitle\abstracts{
The role of the dense matter properties on the tidal deformability and gravitational waveforms of binary neutron stars is studied using a set of unified equations of state. Based on the nuclear energy-density functional theory, these equations of state provide a thermodynamically consistent treatment of all regions of the stars and were calculated using functionals that were precision fitted to experimental and theoretical nuclear data.}

\section{Introduction}

Observations of gravitational waves (GW) from binary neutron-star (NS) mergers offer a new way to probe the interior of these stars, supposed to contain distinct regions: 
an outer crust made of a Coulomb crystal of nuclei with degenerate electrons, an inner crust where nuclear clusters coexist with free neutrons and electrons, and a liquid core (see, e.g. Ref.~\cite{bc2018} for a review).

\section{Unified equations of state for neutron stars}

\subsection{Brussels-Montreal energy-density functionals}

\begin{figure}
\begin{minipage}{0.4\linewidth}
\centerline{\includegraphics[width=1.1\linewidth]{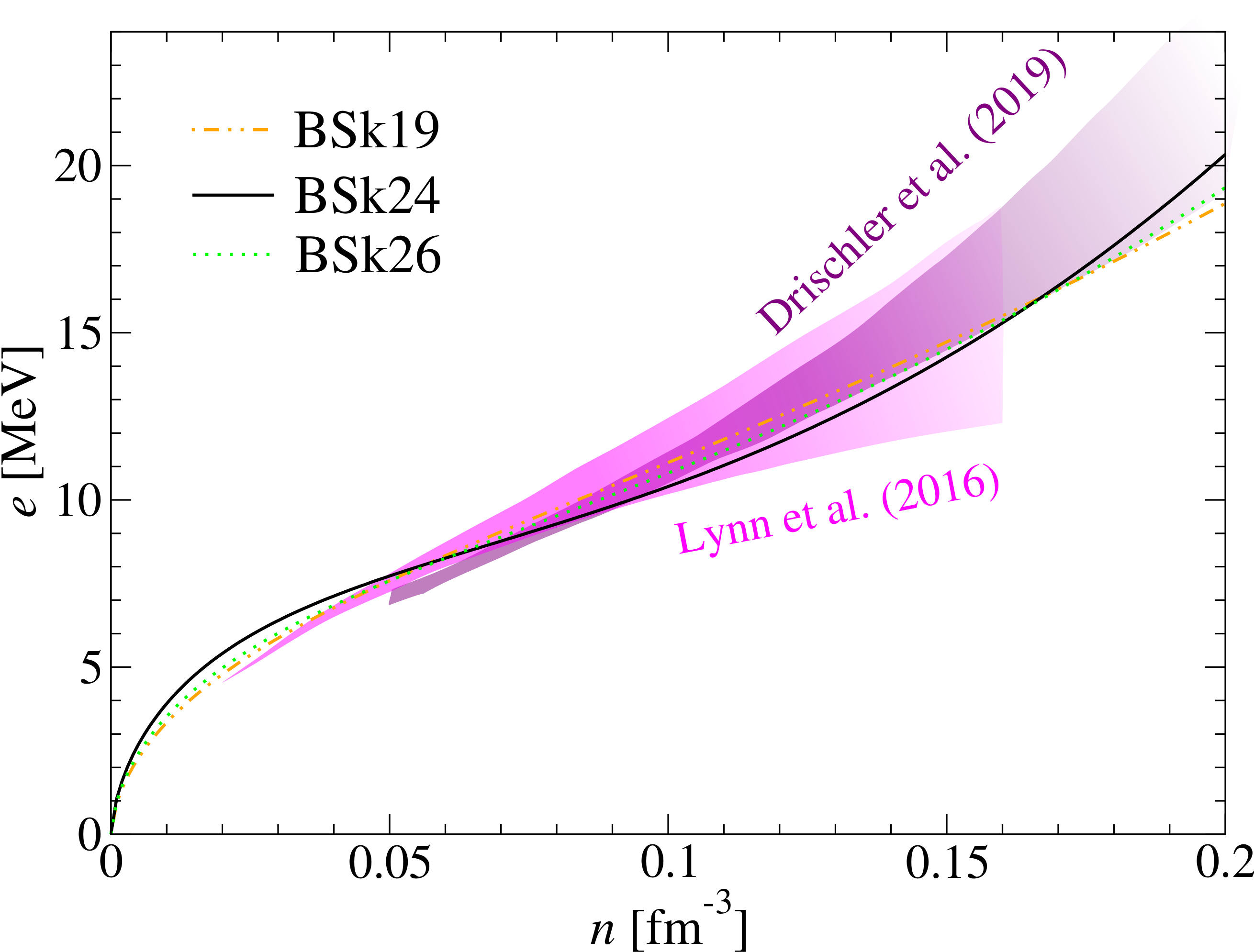}}
\end{minipage}
\hfill
\begin{minipage}{0.4\linewidth}
\centerline{\includegraphics[width=1.1\linewidth]{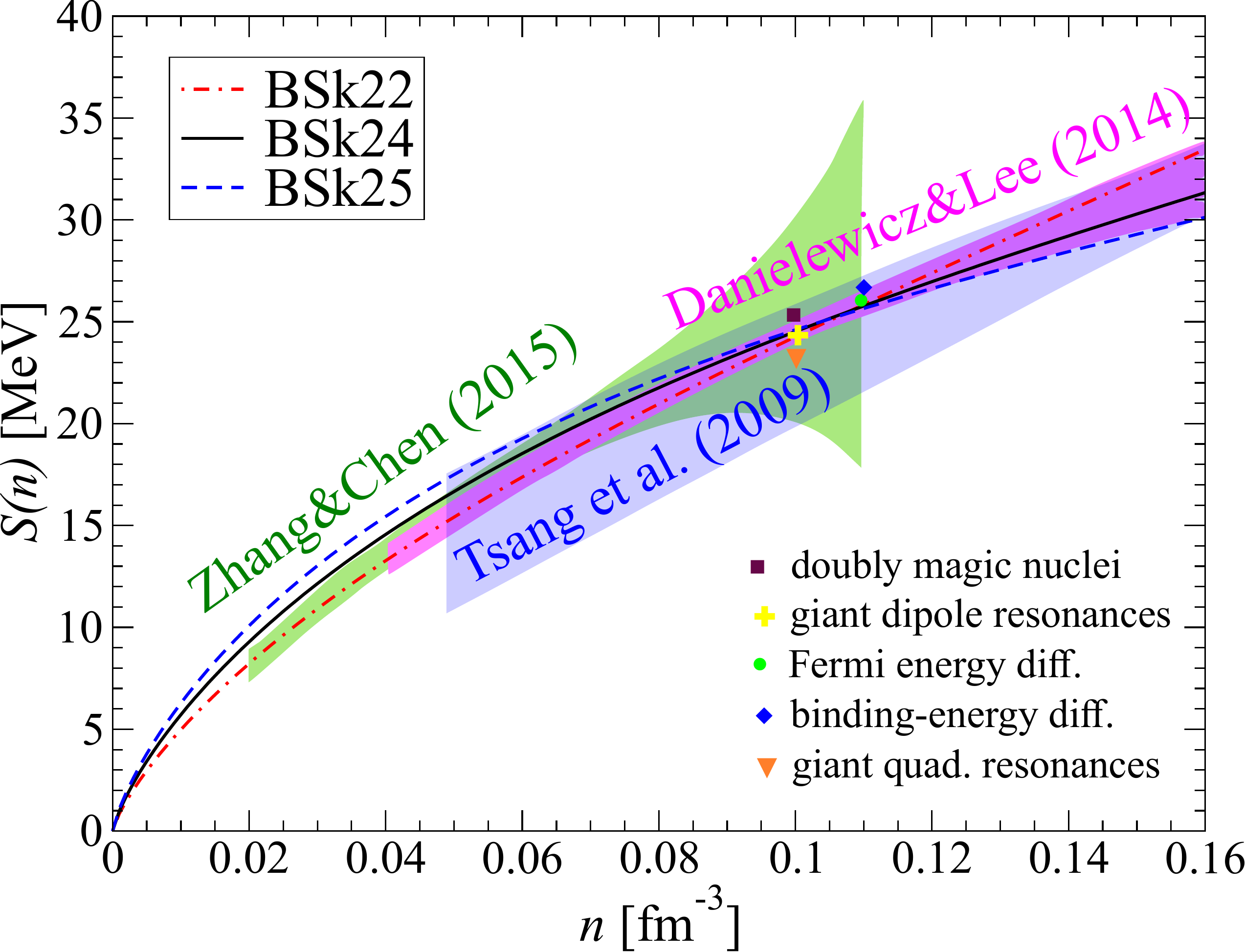}}
\end{minipage}
\caption[]{\textbf{Left:} Energy per particle in neutron matter vs density $n$. The shaded areas represent constraints obtained from chiral effective field theory~\cite{lynn2016,drischler2019}.
\textbf{Right:} Symmetry energy. The shaded areas are experimental constraints: from heavy-ion collisions~\cite{tsang2009} (blue), isobaric-analog states and neutron skins~\cite{danielewicz2014} (purple), and the electric dipole polarizability of $^{208}$Pb~\cite{zhang2015} (green). See Ref.~\cite{perot2019} and references therein for details.}
\label{fig:eos}
\end{figure}

In the nuclear energy-density functional (EDF) theory~\cite{bhr03}, nucleons are treated as independent quasiparticles in a self-consistent Hamiltonian via the Hartree-Fock-Bogolyubov (HFB) method. A family of EDFs~\cite{gcp10,gcp13} have been recently constructed, all based on extended Skyrme 
effective interactions~\cite{cgp09} and precision-fitted to essentially all experimental atomic mass data with a root-mean square deviation as low as 0.5-0.6 MeV.
To assess the role of nuclear uncertainties, BSk19, BSk24 and BSk26 were simultaneously fitted to different realistic neutron-matter equations of state (EoS) with different degrees of stiffness, while BSk22, BSk24 and BSk25 mainly differ in their predictions for the symmetry energy, as shown in Fig. \ref{fig:eos}.  

\subsection{Consistent description of the different regions of a neutron star}

The interior of a NS is supposed to be cold and fully catalysed. A detailed account of the calculations in the different regions can be found in Ref.~\cite{pearson2018}. The properties of the outer crust for densities $\rho \gtrsim 10^6$~g~cm$^{-3}$ were determined using experimental atomic mass data supplemented by HFB predictions. For the inner crust, the 4th-order extended Thomas-Fermi method was adopted within the Wigner-Seitz approximation using parametrised nucleon distributions. Proton shell and pairing corrections were added perturbatively. The core was assumed to contain nucleons and leptons.
Complete numerical results and fits can be found in Refs.~\cite{pearson2018,potekhin2013}.

\section{Tidal deformability of neutron stars}

\subsection{Definitions}

In a binary system, each NS is deformed due to tidal forces. The tidal field can be decomposed into an ``electric'' component $\mathcal{E}_L$, where $L$ denotes a set of space indices $i_1 i_2\dotsi i_\ell$, and a ``magnetic'' component $\mathcal{M}_L$ (absent in Newtonian theory), inducing inside the star a mass multipole moment $\mathcal{Q}_L$ and a current multipole moment $\mathcal{S}_L$, given to leading order by 
\begin{equation*}
\mathcal{Q}_L = \lambda_\ell\,\mathcal{E}_L\, , \hspace{2cm}
\mathcal{S}_L = \sigma_\ell\,\mathcal{M}_L \, ,
\end{equation*}
where $\lambda_\ell$ and $\sigma_\ell$ are respectively referred to as the gravitoelectric and gravitomagnetic tidal deformabilities of order $\ell$, and are related to the corresponding Love numbers through:
\begin{equation*}
k_\ell = \frac{1}{2}(2\ell-1)!!\,\frac{G\lambda_\ell}{R^{2\ell+1}}\, , \hspace{2cm}
j_\ell = 4(2\ell-1)!!\,\frac{G\sigma_\ell}{R^{2\ell+1}} \, ,
\end{equation*}
where $G$ is the gravitational constant. The circumferential radius $R$ of the star and the Love numbers depend on the dense-matter EoS. The formalism to calculate Love numbers can be found in Ref.~\cite{perot2021} with numerical results up to $\ell=5$. 
The measurable dimensionless tidal deformability coefficients are given by 
 \begin{equation*}
    \Lambda_\ell = \frac{2}{(2\ell-1)!!} k_\ell \Big(\frac{c^2 R}{G M}\Big)^{2\ell+1}\, , \hspace{2cm}
    \Sigma_\ell = \frac{1}{4(2\ell-1)!!} j_\ell \Big(\frac{c^2 R}{G M}\Big)^{2\ell+1}\, ,
\end{equation*} 
with $M$ the gravitational mass and $c$ the speed of light.

\subsection{Role of dense matter on tidal deformability and gravitational-wave signal}

\begin{figure}[t!]
\begin{minipage}{0.4\linewidth}
\centerline{\includegraphics[width=1.15\linewidth]{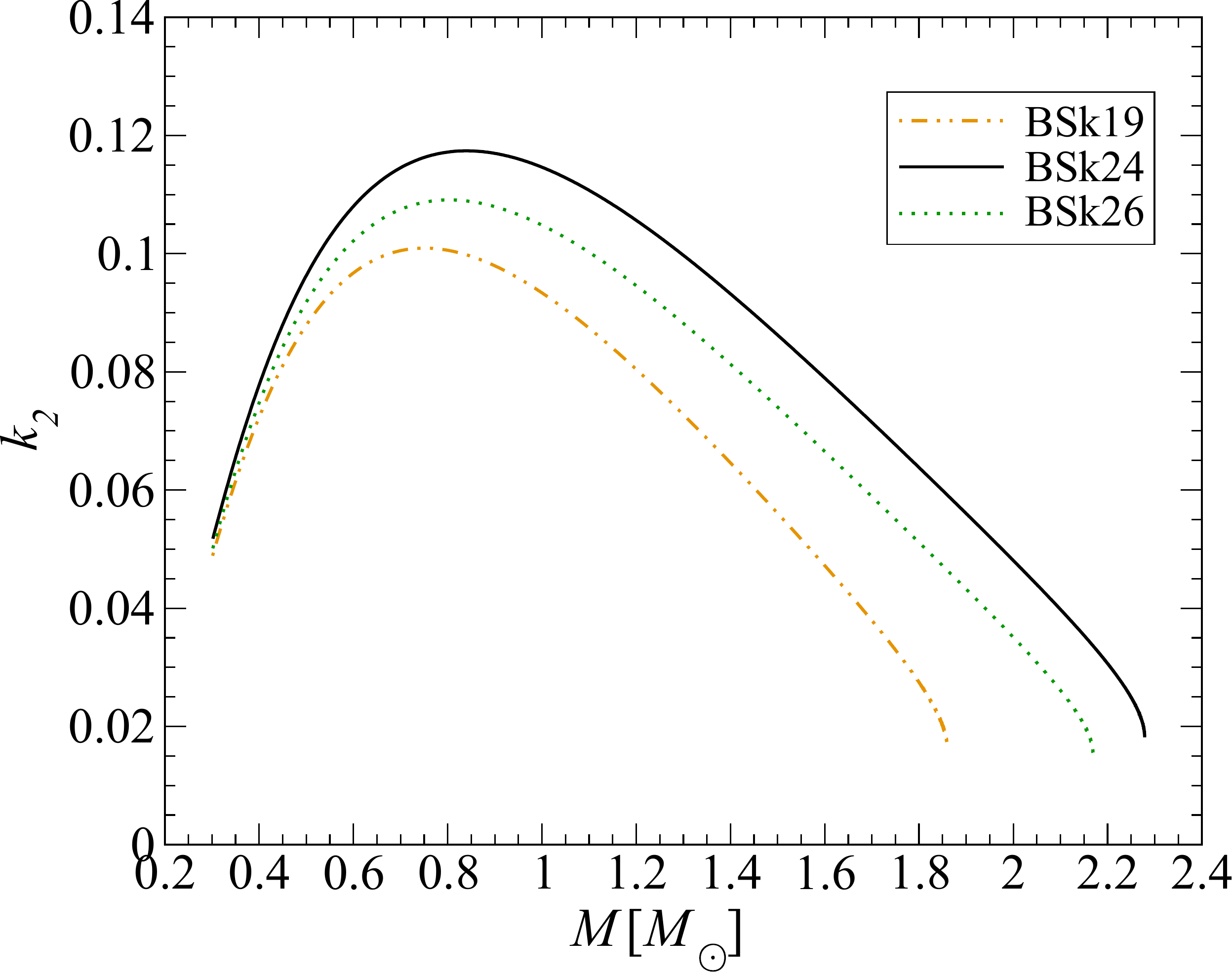}}
\end{minipage}
\hfill
\begin{minipage}{0.4\linewidth}
\centerline{\includegraphics[width=1.15\linewidth]{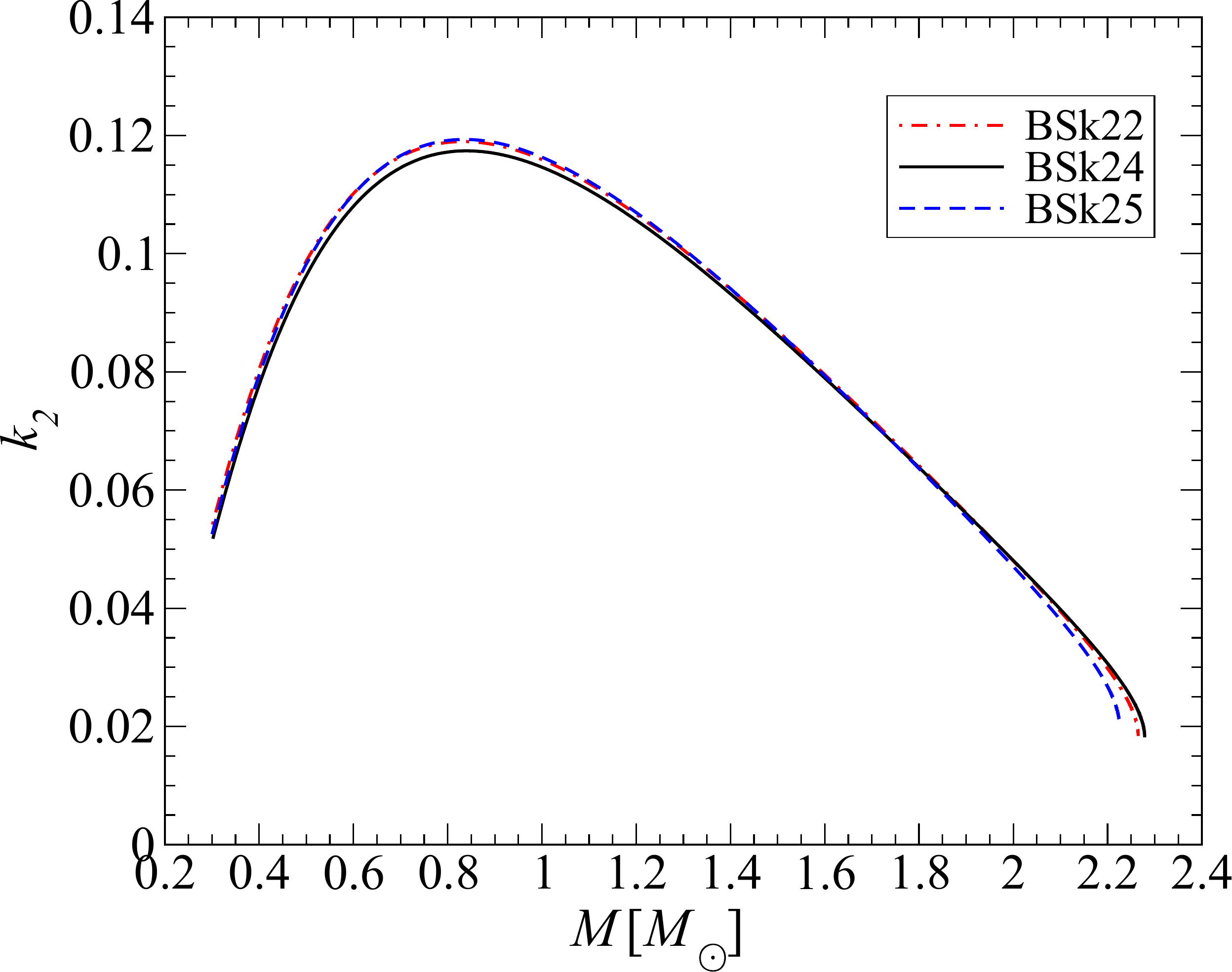}}
\end{minipage}
\caption{Second gravitoelectric Love number $k_2$ as a function of the NS mass $M$.}
\label{fig:love}
\end{figure}

Comparing results for BSk22, BSk24 and BSk25 shows that the symmetry energy plays a minor role both for $k_\ell$ and $j_\ell$ (see Fig. \ref{fig:love} for $k_2$).
The key factor appears to be the stiffness of the neutron-matter EoS, as can be seen by comparing results for BSk19, BSk26 and BSk24 (by increasing order of stiffness): the softer the EoS is, the lower is the value for $k_\ell$ and $j_\ell$ (in absolute value) for a given NS mass.
The impact of the symmetry energy on $k_\ell$ becomes more visible with increasing $\ell$, contrary to $j_\ell$. Full results up to $\ell =5$ are presented in Ref.~\cite{perot2021}. 
The magnitude of both the gravitoelectric and gravitomagnetic Love numbers decreases with increasing $\ell$. For a $1.4 M_\odot$ NS with the BSk24 EoS, the values of $k_5$ and $j_5$ represent only 4\% and 3\% of those of $k_2$ and $j_2$, respectively. 
The Love numbers $k_\ell$ and $j_\ell$ are found to be very sensitive to the EoS of the crust, but mainly through the stellar radius~\cite{perot2020}. On the other hand, the crust has a negligible impact on the measurable tidal deformability coefficients $\Lambda_\ell$ and $\Sigma_\ell$. 

\begin{figure}[t!]
\begin{minipage}{0.4\linewidth}
\centerline{\includegraphics[width=1.28\linewidth]{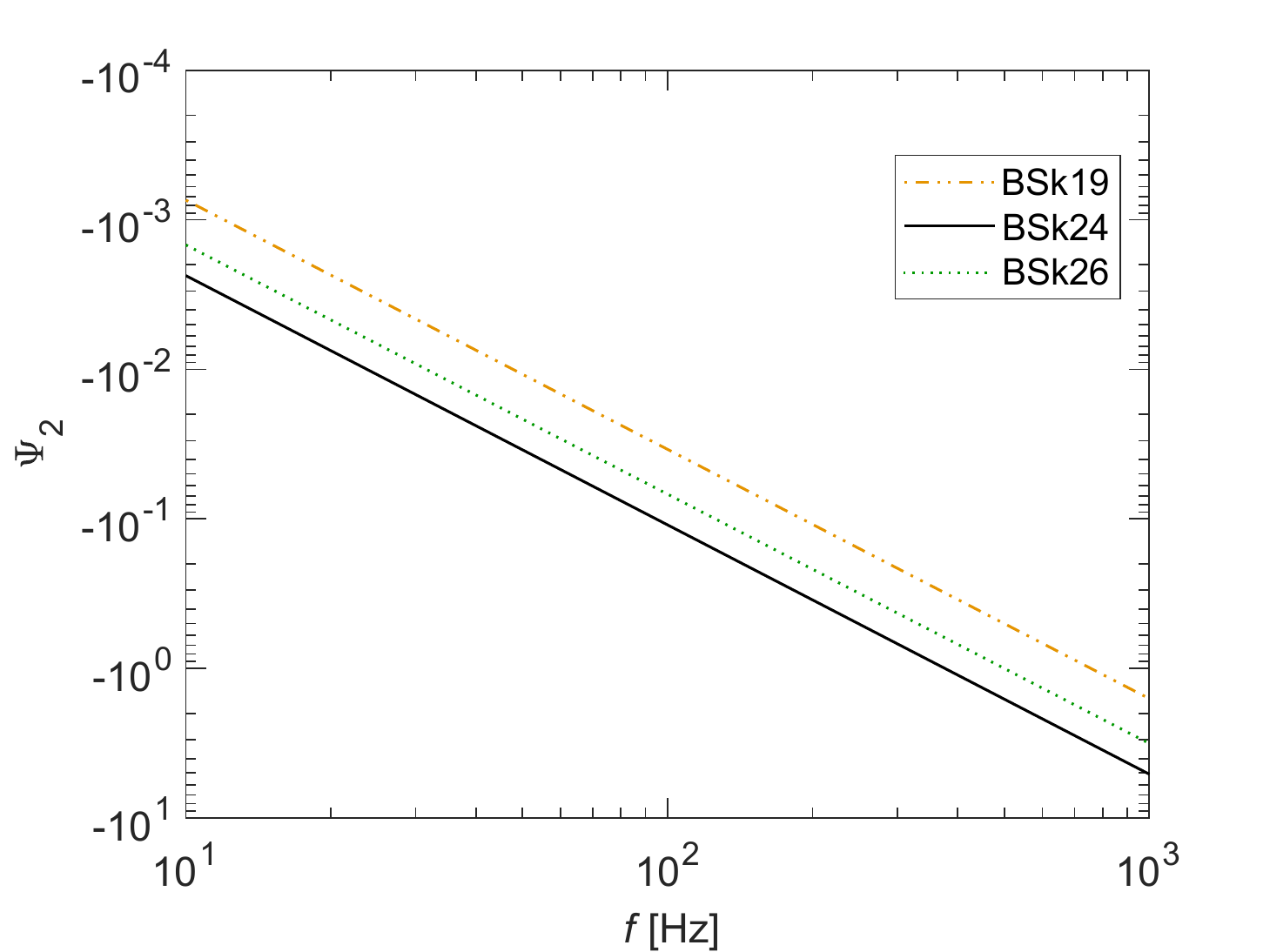}}
\end{minipage}
\hfill
\begin{minipage}{0.4\linewidth}
\centerline{\includegraphics[width=1.28\linewidth]{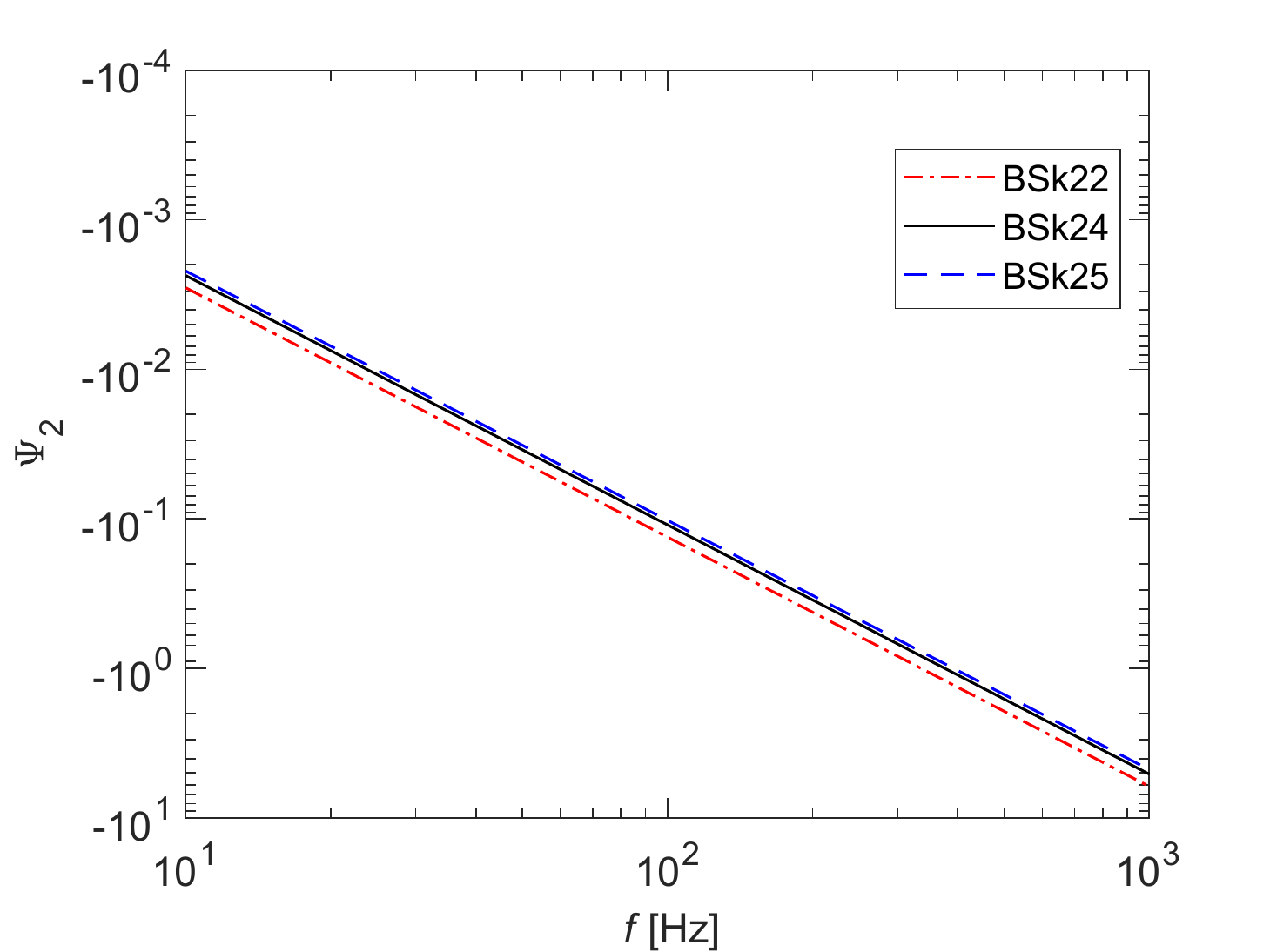}}
\end{minipage}
\caption{Tidal correction to the GW phase associated with the second gravitoelectric Love number $k_2$, as a function of the frequency $f$ and considering binary systems with both NSs having a mass $M=1.4 M_\odot$.}
\label{fig:GW-phase}
\end{figure}

Extracting information about the symmetry energy from the GW signal during the inspiral phase will be very difficult, as can be seen by comparing results obtained for BSk22, BSk24, and BSk25 in the right panel of Fig. \ref{fig:GW-phase}. On the other hand, the stiffness of the neutron-matter EoS leaves a clear imprint on the waveform as shown in the left panel of Fig. \ref{fig:GW-phase}. The comparison of the results obtained for BSk19, BSk26 and BSk24 shows that the softer the EoS is, the more pronounced are the tidal effects. 
The relative importance of the different $\ell$-terms is found to follow the same hierarchy as the Love numbers.
For detailed results, see Ref.~\cite{perot2021}. 

\subsection{Observational constraints}

\begin{figure}[t!]
\begin{minipage}{0.4\linewidth}
\centerline{\includegraphics[width=1.15\linewidth]{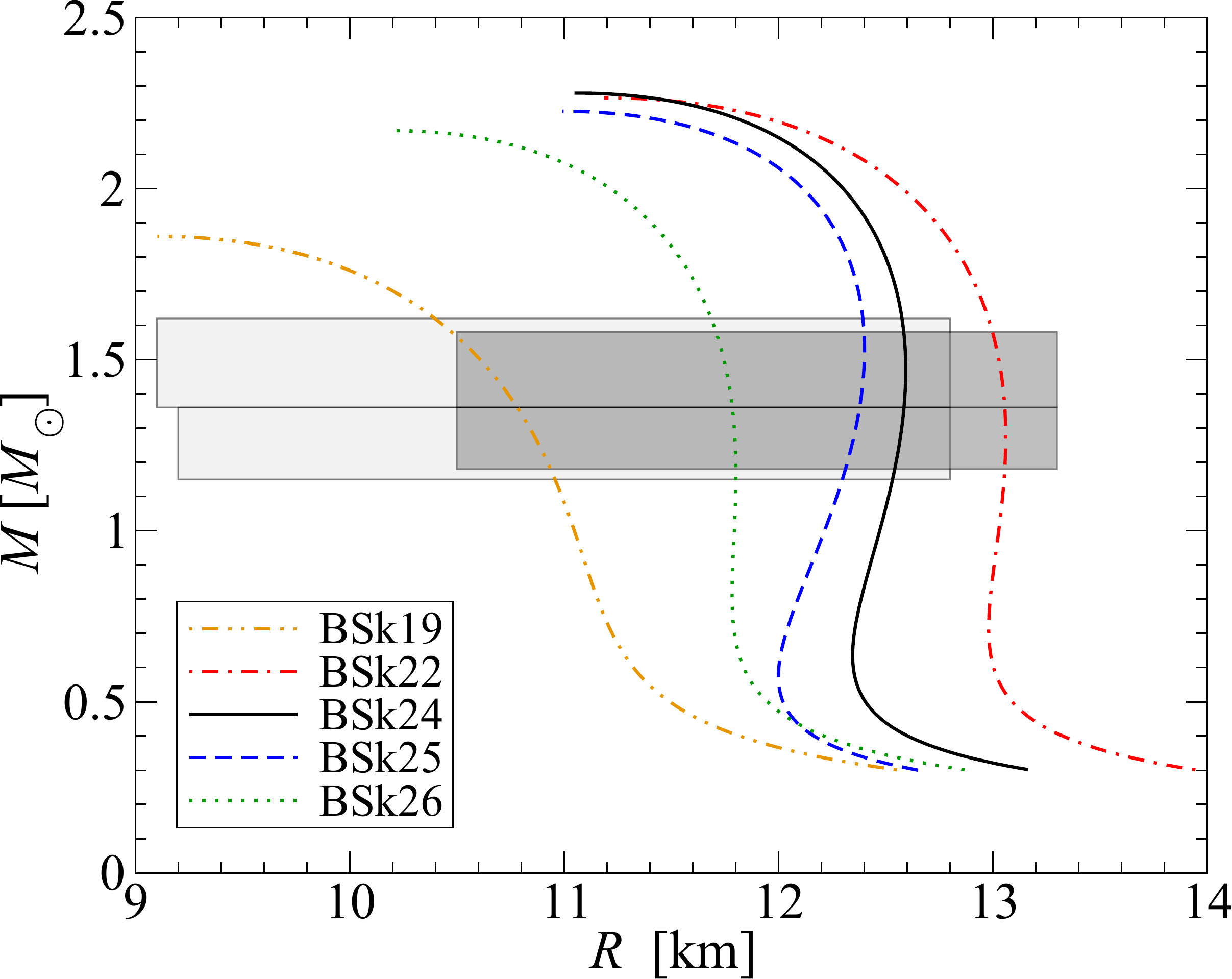}}
\end{minipage}
\hfill
\begin{minipage}{0.4\linewidth}
\centerline{\includegraphics[width=1.15\linewidth]{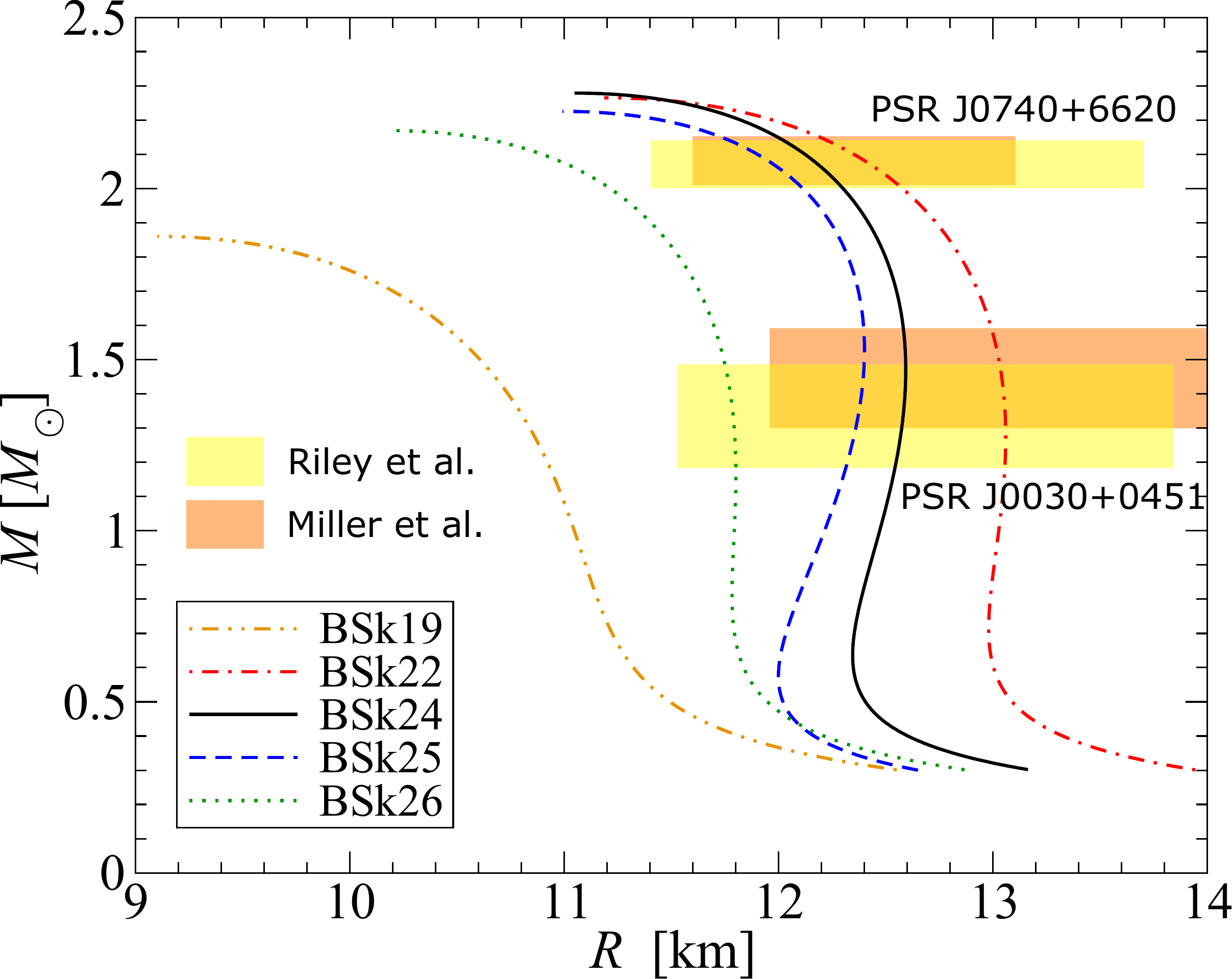}}
\end{minipage}
\caption[]{Mass-radius relations of NS for the BSk EoS. \textbf{Left:} The boxes represent measurements by LIGO-Virgo collaboration from observations of GW170817 using two different methods~\cite{ligo-virgo2018}. \textbf{Right:} The boxes represent measurements of two pulsars from NICER observations~\cite{riley2021,miller2021,riley2019,miller2019}.}
\label{fig:M-R}
\end{figure}

As shown in Fig.~\ref{fig:M-R}, observations of GW170817 by the LIGO-Virgo collaboration~\cite{ligo-virgo2018} do not provide very stringent constraints on the EoS. Our softest EoS for BSk19 remains compatible even though it fails to predict the existence of the most massive NSs, such as PSR~J0740+6620 recently observed by NICER~\cite{riley2021,miller2021}. These latter observations also rule out BSk26~\cite{riley2019,miller2019}.

\section{Conclusion}

The role of dense matter on the tidal deformability of NSs and on the GW signal from binary NSs has been studied using the series of unified EoSs BSk19-26 all based on the nuclear EDF theory. For these EoSs, we have shown that the symmetry energy has essentially no impact on the Love numbers, characterizing the tidal response of NSs, as well as on the tidal phase corrections to the GW signal during inspiral; the key factor appears to be the neutron-matter stiffness: 
the softer the EoS is, the more pronounced are the tidal effects. Moreover, the GW signal appears to be essentially independent of the crust. 

\section*{Acknowledgments}

This work was financially supported by Fonds de la Recherche Scientifique (Belgium) under Grant No. PDR T.004320 and the European Cooperation in Science and Technology Action (EU) CA16214. L. P. is a FRIA grantee of the Fonds de la Recherche Scientifique (Belgium).


\section*{References}

\begin{thebibliography}{99}

\bibitem{bc2018} D. Blaschke and N. Chamel, edited by L. Rezzolla et al.,
Astrophysics and Space Science Library Vol. {\bf 457} (Springer, Berlin, 2018) p. 337-400.
\bibitem{bhr03} M. Bender, P.-H. Heenen and P.-G. Reinhard, Rev. Mod. Phys.{\bf 75}, 121 (2003).
\bibitem{gcp10}S. Goriely, N. Chamel, and J.~M. Pearson,
Phys. Rev. C {\bf 82}, 035804 (2010).
\bibitem{gcp13} S. Goriely, N. Chamel, and J. M. Pearson, Phys. Rev. C {\bf 88}, 024308 (2013).
\bibitem{cgp09} N. Chamel, S. Goriely, and J.~M. Pearson, Phys. Rev. C {\bf 80},
065804 (2009).  
\bibitem{lynn2016} J. E. Lynn et al., Phys. Rev. Lett. {\bf 116}, 062501 (2016).
\bibitem{drischler2019} C. Drischler, K. Hebeler, and A. Schwenk, Phys. Rev. Lett. {\bf 122}, 042501 (2019).
\bibitem{tsang2009} M. B. Tsang et al., Phys. Rev. Lett. {\bf 102}, 122701 (2009).
\bibitem{danielewicz2014} P. Danielewicz and J. Lee, Nucl. Phys. {\bf A 922}, 1 (2014).
\bibitem{zhang2015} Z. Zhang and L.-W. Chen, Phys. Rev. C {\bf 92}, 031301(R) (2015).
\bibitem{perot2019} L. Perot, N. Chamel, A. Sourie, Phys. Rev. C {\bf 100}, 035801 (2019).
\bibitem{pearson2018} J. M. Pearson et al., MNRAS {\bf 481}, 2994 (2018).
\bibitem{potekhin2013} A. Y. Potekhin et al., A\&A {\bf 560}, A48 (2013).
\bibitem{perot2021} L. Perot, N. Chamel, Phys. Rev. C {\bf 103}, 025801 (2021).
\bibitem{perot2020} L. Perot, N. Chamel, A. Sourie, Phys. Rev. C {\bf 101}, 015806 (2020).
\bibitem{ligo-virgo2018} B. P. Abbott et al., Phys. Rev. Lett. {\bf 121}, 161101 (2018). 
\bibitem{riley2021} T. E. Riley et al., ApJL {\bf 918}, L27 (2021). 
\bibitem{miller2021} M. C. Miller et al., ApJL {\bf 918}, L28 (2021). 
\bibitem{riley2019} T. E. Riley et al., ApJL {\bf 887}, L21 (2019).
\bibitem{miller2019} M. C. Miller et al., ApJL {\bf 887}, L24 (2019).
 
\end{thebibliography}
\end{document}